\documentclass{article}
\usepackage{graphicx}
\usepackage{amsmath}
\usepackage{amsfonts}
\usepackage{amssymb}
\newtheorem{theorem}{Theorem}

\begin{document}

\title{An anthology of non-local QFT and QFT on noncommutative spacetime\\{\small Dedicated to Detlev Buchholz on the occasion of his 60th birthday}}
\author{Bert Schroer\\presently: CBPF, Rua Dr. Xavier Sigaud 150 \\22290-180 Rio de Janeiro, Brazil\\Prof. em., Institut f\"{u}r Theoretische Physik, FU-Berlin\\email: schroer@cbpf.br}
\date{May 6, 2004}
\maketitle
\begin{abstract}
Ever since the appearance of renormalization theory there have been several
differently motivated attempts at non-localized (in the sense of not generated
by point-like fields) relativistic particle theories, the most recent one
being at QFT on non-commutative Minkowski spacetime. The often conceptually
uncritical and historically forgetful contemporary approach to these problems
calls for a critical review in the light of previous results on this subject.
\end{abstract}

\section{History of attempts at non-local fields}

To attribute in-depth investigations of non-local QFT to the last decade, as
it is done in most contemporary articles on this subject, is historically
incorrect. There was already a flurry of interest (quite a strong one
considering the number of particle physicist at that time) in non-local
aspects of QFT leading to significant results as far back as the late
50s\footnote{This is of course ment relative to the much smaller number of
particle physicists at that time.}. This interest originated in the wake of
renormalization of QED and of the LSZ scattering theory. The main physical
motivating idea was to get to more and ''better''\ (in the sense of less
singular) interactions by using L-invariant structure functions in the
interaction part of the Lagrangian. In addition there was the desire to
understand whether (pre-QCD) meson-nucleon structure, as it became
experimentally accessible in nucleon formfactors, could be consistently used
in a Lagrangian formalism; in other words if what nowadays would be considered
as ''effective interactions''\ could be of a more fundamental significance
within a non-local Lagrangian setting.

The extensive work of Kristensen and M\o ller \cite{K-M} as well as an
important contribution by Claude Bloch \cite{Bloch} and Hayashi \cite{Hayashi}
with prior remarks by Pauli and Rayski led to lively discussions and also
called the critical attention of leading field theorists of the post
renormalization era as Kallen and Lehmann \cite{K-L}. The issue of
non-locality attracted many Japanese researchers and became even the main
topics of a 1954 conference in Japan \cite{conf}. Since relativistically
invariant non-local interactions even on a formal level do not fit well with
the canonical equal time formalism (and neither with the associated functional
integral approach), most investigation were carried out in the Yang-Feldman
setting which is independent of the canonical formulation and most closely
related to both the Wightman theory and the LSZ framework of scattering
theory. The main problem in this setting, which relates the interacting
Heisenberg fields with on-shell in- and out- fields, is the proof that if the
incoming on-shell field is assumed to have the standard free field commutation
relation then this property is inherited by the on-shell outgoing field, so
that the S-matrix is unitary. This is a problem which is in principle
decidable in perturbation theory. The original structural argument in favor of
this property in C. Bloch's work was not correct\footnote{This can easily be
seen by realizing that e.g. the very existence of creation/annihilation
operators fulfilling the Zamododchikov-Faddeev relations provides a
counterexample.}; indeed it was noticed somewhat later \cite{Hayashi} by
explicit calculation up to fourth order of the commutator of the outgoing
field within the Kristensen-M\o ller setting (which uses a L-invariant
formfactor instead of a pointlike interaction vertex) that it is definitely violated.

Although this was the result of a particular non-local model, it was believed
that this negative answer is generic to non-local interactions. A very clearly
written re-investigation confirming this negative result within a more modern
context appeared two decades later \cite{Marnelius}. Since the standard
derivation of the unitary S-matrix in QFT (as envisaged by Lehmann, Symanzik
and Zimmermann and proven by Haag and Ruelle \cite{Haag}), requires the
validity of the spacelike cluster property for the correlation functions of
the Heisenberg fields, a proof that the on-shell outgoing field does not
fulfill the free field commutation relation shows at the same time that the
cluster property is violated. The latter is part of what one summarily calls
''macro-causality''\footnote{Strictly speaking this is ''macrolocality''
whereas macrocausality is more related to the absence of timelike
precursors.}. Consider a partition of the field coordinates in a correlation
function into two clusters which are separated in the sense that the
localization points of the fields in one cluster are say to the left of those
of the other cluster. Increasing the cluster distance to spacelike infinity,
the vacuum expectation factorizes in those of the two clusters and the
connected goes to zero with the well-known Yukawa fall-off which is given by
the smallest invariant mass of the intermediate states above the vacuum
between the two clusters. However if the clusters inside the expectation value
are overlapping, one first has to disentangle them to the previous
non-overlapping position, which requires the application of spacelike
commutation relations. If the latter are local, there will be no modification
of the fall-off, but the situation may change in case of non-local spacelike commutators.

There exists the following remarkable theorem of Borchers and Pohlmeyer
\cite{B-P} which for Bose fields in a theory with mass gap amounts to the
following statement

\begin{theorem}
A spacelike fall-off of the commutator of the (Heisenberg) fields in the
relative distance which is faster than Yukawa's exponential decrease leads
back to a micro-causal theory.
\end{theorem}

Here faster than Yukawa means that the decrease is dominated by $exp-\lambda
r^{1+\varepsilon}$ with $\lambda,\varepsilon>0$ and $r$ the relative spatial
distance. From this follows that the connected parts of correlation functions
for genuine non-local Poincar\'{e} covariant theories with positive energy
have a weaker large distance decrease than that determined by the standard
intermediate state argument. In addition the clustering decrease is
anisotropic in the sense that it is still Yukawa like for nonoverlapping
clusters and weaker than Yukawa otherwise. Such a behavior is quite unusual in
particle physics; apart from jeopardizing the standard analytic connection
with the Euclidean formulation it may have visible experimental
manifestations, but we will not pursue this interesting problem in this work.

This also indicates that the clustering in genuine non-local QFTs is not any
more determined by the ``kinematical'' properties of the lowest invariant
energy of intermediate states\footnote{The pre-exponential strength factors in
local QFT do of course depend on details of the interaction.} but rather
suffers modifications which depend via the commutator fall-off on the details
of the non-local interaction. In a way these results suggest that, contrary to
what one may have expected intuitively and contrary to the deceivingly
lighthearted manner in which computational tricks as cutoffs and regulators
are sometimes interpreted as basic physical concepts, \textit{causality is an
extremely rugged property} and that a notion of ``a little bit non-local''\ or
``a little bit acausal''\ is not much more sensible than ``a little bit
pregnant''. Whereas it is easy to cut-off or regularize Feynman integrals as
part of perturbative renormalization theory it, is generally not known how
such ideas can be consistently implemented in the operational setting of
Hilbert spaces and operators.

Hence it should not come as a surprise that most attempts of introducing
deviations from micro-causality actually amount to violating macro-causality
and often also Lorentz invariance in the wake; but macro-causality is the
absolute borderline between physics and the realm of poltergeists, and
sacrificing the relativity principle without getting a conceptual benefit does
not seem to hold much attraction either.

Among the more prominent attempts there was the proposal by Lee and Wick to
modify the Feynman rules by pair of complex poles together with their complex
conjugates. After formulating this idea in a field theoretic setting in such a
way that at least the unitarity of the S-matrix within the Yang-Feldman
setting was maintained, Marques and Swieca \cite{M-S} showed that the proposal
led to physically untenable timelike power ''precursors''\footnote{Only in the
case of external interactions only the precursors have a more amenable
exponential behavior.}.

This raised the question whether there exist consistent relativistic unitary
and macro-causal particle theories at all. A positive answer was given
Coester's idea \cite{Coester} of construction of ''direct particle
interactions''. This was first formulated on 3-particle systems and then
generalized (in collaboration with Polyzou) \cite{C-P} to multiparticle
theories. As a pure relativistic particle theory without vacuum polarization,
it turns out to have no natural second quantization setting; but on the other
hand it fulfills all properties which are expressible in terms of particles
without inferring fields. In particular these theories fulfill the cluster
separability properties of the associated Poincar\'{e} invariant unitary
S-matrix; in fact even the generators of the Poincar\'{e} group cluster
(though not with Yukawa like exponential decrease) as a result their existence
contradicts a dictum (ascribed to S.Weinberg) saying that a Poincar\'{e}
invariant unitary S-matrix which clusters is characteristic for local QFT. The
possibility that this conjecture may apply to the stronger requirement of a
Yukawa type clustering as known from local QFT is however not completely excluded.

Although these Coester-Polyzou (C-P) theories of direct particle interactions
have acquired some popularity in phenomenological treatment\footnote{The work
on direct particle interactions has been mainly published in nuclear physics
journals which perhaps explains why it is virtually unknown in the high energy
physics community.} of medium energy meson-nucleon interaction calculations
(where the limitation in energy prevent the appearance of real multi-particle
creation \cite{C-P}), our interest in them, as far as this paper is concerned,
is strictly limited to the desire to broaden our conceptual scope of the
meaning of macrocausal and non-local behavior even if the objects which
achieve that are strictly speaking not QFTs since, as in the case of the C-P
theories, they allow no natural second quantization. As soon as one leaves the
setting of local QFT, properties as vacuum polarization, TCP,
spin\&statistics..which are structural consequences of locality loose their
validity, although one may in many cases add them by hand i.e. select those
non-local models which have these additional properties. One should also
notice that, unlike the Galilei group, the Poincar\'{e} group representations
do not impose any mass superselection rule and hence one can incorporate
particle creation processes into a direct particle interaction setting.

Besides these theories which, as a result of absence of a second quantization
setting are not QFTs in the strict sense, no consistent non-local but
macrocausal relativistic particle theories are presently known. The main
subject of this paper will consists in presenting some recent attempts of how
to go beyond locality in ``noncommutative QFT''; more specifically how to
achieve this by formulating standard QFT on noncommutative Minkowski spacetime.

With the above mentioned negative results coming from the history of non-local
QFT from different sources, the studies about the feasibility of Lorentz
invariant non-locality was laid to rest up to the beginnings of the 90's, when
quite different well-founded physical considerations led to ideas about
quantum field theory on non-commutative Minkowski spacetime. These turned out
to have a more subtle relation to the issue of non-locality. The starting
point was an observation by Doplicher, Fredenhagen and Roberts \cite{D-F-R} on
a quasiclassical interpretation of Einstein's general relativity equation with
the added requirement that mere localization measurements (without the action
of additional dynamical processes) should not generate black hole horizons.

Such a requirement was not entirely new; one finds a mentioning of
localization measurements and black holes within a very restrictive (and
physically somewhat less motivated) setting as far back as 1964 \cite{Mead}.
However a formulation of uncertainty relations allowing for localization
regions of any shape, derived from a few physically plausible assumptions,
really starts with the DFR work.

Nowadays one occasionally also finds references to a much older 1947 paper by
Snyder \cite{Snyder}. This paper is sometimes quoted in a slightly wrong
context since Snyder's motivating idea was to mimic lattice regularization in
a covariant way, which for technical reasons led him to noncommuting spatial
position operators. Without the clarification of the idea of renormalization
there was anyhow no chance in 1947 to say something relevant on quantum field
theoretic aspects of gravitation.

The DFR argument in favor of uncertainty relations for position operators is
reminiscent of the Bohr-Rosenfeld derivation of uncertainty relations for the
electromagnetic field based on a quasiclassical interpretation of
electromagnetism coupled to Schroedinger quantum matter. These uncertainty
relation still leave many possibilities for commuting relations between
coordinate spacetime operators and DFR have chosen the simplest one. The
terminology ``non-commutative Minkowski spacetime'' refers to the fact that
even in the absence of curvature the DFR model describes a non-commutative
modification of Minkowski spacetime. The Wigner's particle picture is fully
incorporated in the DFR setting,\ but the Fourier transform which for standard
free fields carries the canonical Wigner momentum space creation/annihilation
operators to the spacetime dependent fields involves now operators $q_{\mu}$
obeying commutation relations which are described in terms of an antisymmetric
matrix $Q$ with variable center-valued entries.

It is not clear how the original motivation of DFR of reconciling general
relativity with quantum physics can be implemented since the uncertainty
relations as well as the commutation relations have not been reconciled with
the property of general covariance and background independence. Therefore the
DFR non-commutative setting is presently limited to explore the concept of QFT
on non-commutative Minkowski spacetime.

The novel feature of QFT on non-commutative spacetime is that in addition to
the Wigner multi-particle momentum Fock space there is an additional Hilbert
space of wave functions which describes the localization aspects in $q$-space
and on which the Lorentz transformations also act. Picking a family of
localizing wave functions, e.g. the family which belongs to minimal
localization (which cannot be pointlike in all 4 components as a result of the
uncertainty relations \cite{D-F-R}) and forgetting that these wave functions
(unlike test functions) suffer an active change under L-transformations as ,
this setting may appears like a modern form of the pre-Einsteinian
''ether''\footnote{Indeed in most treatments the center-valued $Q$ are
assigned fixed numerical values and hence the relativity principle in form of
the 10-parametric family of inertial systems is violated in terms of the
return of an ether.}. But in a setting where the center-valued $Q$ remain
dynamical variables in an altogether Poincar\'{e} covariant representation
space, this historical analogy may be is misleading.

What seems to be implied in the DFR setting is a radical paradigm change of
the meaning of ``relativity principle'' and ``inertial systems'' in a
direction of which the conceptual basis has not been completely understood.
Only if for some reason we were to work in an irreducible representation of
the coordinate operators (in which their commutators obey necessarily
non-L-invariant numerical valued commutations relations) and the Poincar\'{e}
group ceases to have an automorphic action on the algebra, this would indeed
amount to a return of a kind of (quantum) ether. \ One also should refrain
from interpreting the $q^{\prime}s$ as some kind of quantum mechanical
observables (the terminology ``uncertainty relations'' in \cite{D-F-R} may
invite such an interpretation) as e.g. a quantum mechanical Newton-Wigner
\cite{N-W} localization operator\footnote{Although the asymptotically
covariant Newton-Wigner localization does not play any direct role in the
formulation of QFT on non-commutative Minkowski spacetime, it is crucial (as
for any relativistic particle theory) in the formulation of the asymptotic
behaviour which leads to the scattering operator.}.

The conceptual problems posed by such an unusual situation of a QFT on
noncommutative Minkowski spacetime are best discussed in the setting of free
fields where no problems are encountered (see later). However there are
serious difficulties with interactions. Intuitively one would expect that the
noncommutative Minkoski space is an arena for quantum fields which for large
distances presents itself as the ordinary Minkowski spacetime. Since on-shell
quantities like the S-matrix are not directly dependent on short distance
properties (but only indirectly via the encoding into long distances via
particle scattering data) one would think that the very concepts of incoming
and outgoing (Wigner) multi-particle states are remaining intact. However, as
will be seen later, there are serious problems with these expectations about
scattering concepts.

There is a lack of standard covariance, although a less radical one, in the
already mentioned ''direct particle interaction''\ models since the
requirement of additivity in particle interactions and the cluster requirement
cannot be met simultaneously in one formula for the Lorentz generators. As in
the case of the DFR setting, the physics of the C-P theory depends on
additional data (on localization wave functions in the case of the D-F-R
model, the choice of the scattering equivalence in the C-P setting). This to
be expected because the D-F-R noncommutative localization and the C-P
implementation of cluster properties depend on additional choices.

Starting from some observations about the role of noncommutative tori in
string theory, there has been a different line of thought leading to
noncommutative and explicitly non-L-invariant theories. Such attempts are of
course expected to carry all the conceptual fragility of string theory, which
employs a lot of advanced mathematical techniques but tends to be atrophic on
the side of physical concept and physical interpretations. But even if a
strong commitment to history and faithfulness to the principles underlying
particle physics would make it difficult to accept arguments from string
theory at face value, there always remains the possibility to critically
analyze them (independent of their origin) in the light of the conceptual
framework of QFT. This will be the strategy in the rest of this section.

There exist several papers in which (motivated by string theory) the authors
propose numerical-valued commutators between spacetime coordinates which break
Lorentz invariance down to a subgroup. For example in \cite{A-G} the subgroup
of $P(4)$ turns out to be $P(2)\otimes E(2)$ i.e. tensor product symmetry of a
two-dimensional $P(2)$ Poincar\'{e}- with the two-dimensional Euclidean- group
$E(2).$ The associated locality is that of a 1+1 dimensional QFT of the $P(2)$
factor, whereas the $E(2)$ Euclidean degrees of freedom are just
''spectators''\ as far as locality is concerned.

As a logical consequence of this assumption and the well-known fact that
causality imposes a Lorentz invariant shape on the associated positive energy
momentum spectrum \cite{Haag}, one finds that so that the 1+3-dimensional
Wigner particle structure gets lost and with it the relevance of such a model
for 1+3-dimensional particle physics. As a 1+1-dimensional local theory it
naturally fulfills the associated TCP theorem, which combined with the
reflection invariance of the Euclidean spectator theory then leads to a higher
dimensional TCP operation\footnote{In the same vein the analytic spacetime
properties of the model are fixed by the Bargmann-Hall-Wightman theorem
\cite{S-W} in d=1+1.}. In this connection one should recall that QFT,
different from QM, is conceptually intrinsic. In other words something which
smells, trumpets and appears like an elephant really is an elephant.

Apparently the authors in \cite{Chaichian} tried to address these problems.
They propose a theory which maintains the usual particle concept associated
with the mass hyperboloids, but breaks L-invariance through the application of
a Moyal star-functor to the Wightman functions of an interacting local
Poincar\'{e}-invariant QFT (see equ. II.3 in \cite{Chaichian}, the
mathematical difficulties in attributing a precise meaning are not discussed
there). They show that if one lets this functor also change the commutator to
a kind of spacelike star-commutativity, the weak locality assures that the old
TCP operator and the S-matrix remain unchanged under the action of the star
functor. If instead one were to change the local action by a star-product
prescription, none of structural properties as TCP, spin\&statistics,
Jost-Lehmann-Dyson representation etc. is expected to survive.

The star-functor (as any star functor belonging to a fixed set of
numerical-valued $q$-commutators) does not commute with the Lorentz boosts and
therefore ''breaks''\ L-invariance but without harming the Wigner mass shell
support of particles; in fact it preserves precisely the aforementioned
subgroup and results in the D-F-R setting from choosing minimally localizing
wave functions on noncommutative Minkowski spacetime (which lead precisely to
the numerical values of $Q_{\mu\nu}$ used in \cite{Chaichian} and other papers
quoted therein). But in order to attribute meaning to this words one should
give an intrinsic physical definition under what circumstances a mathematical
subgroup can be called a ''broken symmetry''\ of the ambient group. Aside from
the well-defined notion of spontaneous symmetry breaking (this happens to
internal symmetry groups in the Goldstone setting and to the Poincar\'{e}
group for a QFT in a thermal KMS state where the boosts fail to be
implementable), I am not aware of such a meaning. It seems to me the
acceptance of such a terminology as a part of particle physics would indeed
amount to a return of the pre-Einsteinian non-dynamic ether in the setting of
quantum physics.

In regards to the $P(2)\otimes E(2)$ symmetry, there is an interesting analogy
with the 7-parametric subgroup of the Poincar\'{e} group which features as the
symmetry group of the lightfront and which contains the 4-parametric
$P(2)\otimes E(2)$ in a natural way \cite{schroer}. The difference is that in
the process of holographic lightfront projection the spacetime interpretation
changes radically in accordance with the fact that the lightfront is not a
hyperbolic manifold. This change is necessary in order to be able to return to
the \ Poincar\'{e} covariant ambient theory. In this case the fact that the
symmetry is reduced has a clear-cut physical meaning in terms of the
holographic spacetime encoding whereas in the interpretation of $P(2)\otimes
E(2)$ within a noncommutative setting the broken symmetry support has no
physical meaning; in particular the Poincar\'{e} symmetry $P(4)$ does not
return for large distances.

There is of course also the general feeling that one should not be able to
generate a new physical reality by doing nothing else than applying a functor
to an existing theory. In this context it is helpful to remind oneself that
the Heisenberg (commutator-) ring is non-isomorphic to the classical Poisson
ring, i.e. the transition from classical to quantum physics is
''artistic''\ and not functorial (remember Nelson's famous saying: second
quantization is a functor but first quantization remains a mystery).

The only way to counteract the return to a (quantum) ``ether'' however is to
find an extension in which that $P(2)\otimes E(2)$ symmetric theory becomes
part of a larger Poincar\'{e}-invariant setting. This is precisely what the
D-F-R theory \cite{B-D-F-P}\cite{Bahns}\cite{Piacitelli} achieves; it does so
by building a QFT over a non-commutative version of fully Poincar\'{e}%
-invariant Minkowski spacetime. In a more colorful artistic terminology one
may say that the static ether is rendered dynamic i.e. by not simply combining
the possible numerical values of $Q_{\mu\nu}$ into a direct integral or fibre
bundle, but semantic wrappings insinuating analogs can be no substitute for a
yet incompletely understood conceptual situation.

Interpreting the word ''non-local''\ in its widest sense as referring to
\textit{any theory whose content cannot be expressed in terms of pointlike
fields\footnote{In the terminology of AQFT any theory whose net of algebras
cannot be generated by additivity from arbitrarily small double cone localized
algebras.}}, the recently discovered string localized fields also feature
under the heading ''non-local''\ in this article. By string-localized fields
we mean fields $A(x,e)$ as they arose in connection with constructing the
field theory behind the massless infinite spin (rather infinite helicity
tower) Wigner representation of the Poincar\'{e}-group \cite{M-S-Y}. Here $x$
denotes a point in Minkowski space and $e$ is a ``fluctuating spacelike
direction''\ i.e. a localization point of a quantum field in 3-dimensional de
Sitter space. Such strings are causal if their spacelike half-lines $x+R_{+}e$
are relatively spacelike (not just their endpoints). There are very good
reasons to believe that there exist interactions of string-localized fields
which maintain the string localization in each order. There also exist
semiinfinite string-localized massive free fields. Even though they generate
the same Wigner one-particle spaces as their pointlike counterparts, there are
good reasons to believe that they widen the concept of particle interactions
while maintaining appropriately adjusted causal localization properties
\cite{Project}.

Finally it is worthwhile to point out that non-local objects play a crucial
intermediate auxiliary role in the nonperturbative construction of local
theories. A well-known example is supplied by the
vacuum-polarization-free-generators (PFGs) of d=1+1 factorizing models whose
Fourier transforms are momentum space creation/annihilation operators which
fulfill the Zamolodchikov-Faddeev algebra relations and are therefore
non-local. They turn out to be wedge-localized; in fact they generate the
wedge algebra. They do not have the form of a smeared pointlike localized
field with the wedge being the support of the smearing function; rather they
are linked in an inexorable way to the global wedge region \cite{B-B-S}. The
only way to get to sharper localized subwedge algebras is by intersecting
wedge algebras. In this process the generators change; they acquire vacuum
polarization and are represented in terms of infinite series in the Z-F operators.

These ideas about non-local generators of wedge algebras are important,
because they show that one of the motivating pillars for studying non-local
theories namely the idea that local theories are inherently beset by
ultraviolet problems is not true. What causes these problems is not the local
structure of QFT but rather the use of \ ''pointlike field coordinatization''
in the calculational approach. Whereas the use of problem-creating singular
coordinates in geometry can be avoided, the ''coordinates'' of standard QFT
obtained via Lagrangian quantization in the form of fields are inherently
singular. This is of course the main reason why the algebraic approach which
is based on nets of (bounded operator) algebras was proposed instead of the
standard pointlike field formulation. There are myriads of (composite) fields
which all generate the same net of algebras. In a way the spacetime indexed
net of algebras corresponds to local equivalence classes of fields (Borchers
classes) and not to individual pointlike fields. Hence the motivation for
using non-local operators, either as auxiliary tools in local theories or for
the construction of inherently non-local theories is rather similar, in both
cases one wants to avoid ultraviolet problems.

In our presentation we avoid anything which could be understood as an
axiomatic formulation. We think that the scarcity of controllable examples and
the poor mathematical and conceptual status presently does not warrant an
axiomatic approach to non-locality and non-commutativity.

The content is organized in the following way.

In the next section we present the relativistic direct particle interaction
setting which fulfills cluster factorization and the time like aspects of
macro-causality (absence of precursors). In section 3 semiinfinite
string-localized fields are introduced, whereas section 4 shows the usefulness
of non-local (in fact wedge-localized) operators in the nonperturbative
construction of local theories. Finally section 5 presents details about QFT
on noncommutative Minkowski spacetime and its encoding into non-local QFT on
standard Minkowski spacetime. Some open problems are mentioned in the last section.

\section{Direct particle interactions and macro-causality}

It has been known since the early days of particle physics that an interacting
relativistic 2-particle system of massive particles (for simplicity of equal
mass) is simply described by going into the c.m. system and modifying the mass
operator in the following way%
\begin{align}
M  &  =2\sqrt{\vec{p}^{2}+m^{2}}+v\\
H  &  =\sqrt{\vec{P}^{2}+M^{2}}\nonumber
\end{align}
The interaction $v$ may be taken as a function of the relative coordinate
which is conjugate to the relative momentum $p$ in the c.m. system; but since
the scheme does not lead to local differential equations, there is nothing to
be gained to insist in such a locality property. One may the follow Bakamjian
and Thomas (BT) \cite{B-T} and choose the $\vec{P}$-generators in such a way
the interaction does not appear\footnote{The group theory alone does not
select a particular way of introducing interactions; in fact the C-P scheme
uses this flexibility in the multiparticle sectors for the implementation of
the all important cluster properties.
\par
{}} in the formulas for the c.m. momentum $\vec{P}=\vec{P}_{0}$, the total
angular momentum $\vec{J}=\vec{J}_{0}$, as well as in the position conjugate
to the c.m. momentum $\vec{X}=\vec{X}_{0},$ whereas the boost generators
depend implicitly on the interaction through%
\begin{equation}
\vec{K}=\frac{1}{2}(HX_{0}+X_{0}H)-\vec{j}\times\vec{P}(M+H)^{-1}%
\end{equation}
The Wigner canonical spin $\vec{j}=\vec{J}_{0}-\vec{X}_{0}\times\vec{P}$
commutes with $\vec{X}_{0}$ and $\vec{P}$ and results from applying the boost
transformation $L_{P}$ on the Pauli-Lubanski vector $W$%
\begin{align}
L_{P}W  &  =\left(  1,0,0,M\vec{j}\right) \\
W_{\mu}  &  =\frac{1}{2}\sum_{\nu\rho\sigma}J^{\rho\sigma}P^{\nu}%
\varepsilon_{\nu\rho\sigma\mu}\nonumber
\end{align}
It is now easy to check that the commutation relations of the Poincar\'{e}
generators are a result of the above definitions and the canonical commutation
relations of the single particle canonical variable which furnish a complete
(irreducible) set of operators in terms of which any operator in the Hilbert
space may be written. Furthermore for short ranged interactions $v$ the
sequence of unitaries $e^{iHt}e^{.iH_{0}t}$ converges strongly towards the
isometric M\o ller operators \cite{Coester}, from which in turn one may
compute the S-matrix%
\begin{align}
\Omega_{\pm}(H,H_{0})  &  =s-\underset{t\rightarrow\pm\infty}{lim}%
e^{iHt}e^{-iH_{0}t}\\
S  &  =\Omega_{+}^{\ast}\Omega_{-}\nonumber
\end{align}
The unitarity of $S$ in the subspace orthogonal to the possible bound states
is a consequence of the identity of ranges of $\Omega_{\pm}.$ Since the
Hamiltonian is frame-dependent, and a relativistic particle theory is only
physically acceptable if the scattering operators are frame-independent, we
must have%
\begin{equation}
\Omega_{\pm}(H,H_{0})=\Omega_{\pm}(M,M_{0})
\end{equation}
which indeed follows from the same kind of short range assumptions
\cite{Coester} which already assured the validity of the asymptotic
convergence. The same short range is also responsible for the cluster
property, namely the statement that the infinite translation of one particle
to spatial infinity results in the S-matrix converging against the identity
(the product of the single particle S-matrices) $\Omega_{\pm}\rightarrow1.$

The BT form of the generators can be achieved for arbitrary number of
particles. As will be seen, the advantage of this form of the representation
is that in passing from n-1 to n-particles the interactions simply add and one
ends up with Poincar\'{e} group generators for an interacting n-particle
system. But whereas this iterative construction in the nonrelativistic setting
complies with cluster separability, this is not the case in the relativistic
context. This problem shows up for the first time in the presence of 3
particles \cite{Coester}. The BT iteration from 2 to 3 particles gives the
3-particle mass operator%
\begin{align}
M  &  =M_{0}+V_{12}+V_{13}+V_{23}+V_{123}\label{additive}\\
V_{ij}  &  =M(ij,k)-M_{0}\nonumber\\
M(12,3)  &  =\sqrt{\vec{p}_{12,3}^{2}+M_{12}^{2}}+\sqrt{\vec{p}_{12,3}%
^{2}+m^{2}}\nonumber
\end{align}
Here $M(12,3)$ denotes the 3-particle invariant mass in case the third
particle is a ``spectator''\ which does not interact with 1 and 2. The
momentum in the last line is the relative momentum between the (12)-cluster
and particle 3 in the joint c.m. system i.e. $\vec{p}_{12,3}=L_{P}^{-1}%
(\vec{p}_{12}-\vec{p}_{3}),\vec{p}_{12}=p_{1}+p_{2}.$ As in the
nonrelativistic case one can always add a totally connected contribution. This
generally changes the S-matrix of the full system while keeping the lower
particle S-matrices. But contrary to the nonrelativistic case the BT
generators constructed with $M$ do not fulfill the cluster separability
requirement. The latter demands that if e.g. the interaction between two
clusters is removed, the unitary representation factorizes into that of the
product of the two clusters. Applied to the case of 3 particles with particle
3 being a spectator, one expects that shifting the third particle to infinity
will result in a factorization $U_{12,3}(\Lambda)\rightarrow U_{12}%
(\Lambda)\otimes U_{3}(\Lambda).$ But what really happens \cite{Wayne} in the
limit is
\begin{equation}
U_{12,3}(\Lambda)\rightarrow U_{1}(\Lambda)\otimes U_{2}(\Lambda)\otimes
U_{3}(\Lambda)
\end{equation}
The reason for this violation of the cluster separability property (as a
simple calculation using the transformation formula from c.m. variables to the
original $p_{i},i=1,2,3$ shows) is that the translation in the original system
(instead of the c.m. system) does remove the third particle to infinity but it
also drives the two-particle mass operator (with which it does not commute)
towards its free value (for an explicit calculation of this limit see
\cite{Wayne}). The BT construction is very well suited to manufacture a
Poincar\'{e} covariant 3-particle interaction which is additive
(\ref{additive}) in the respective c.m. interaction terms, but the
$U(\Lambda)$ of the resulting system will not be cluster-separable in the
sense of the previously two-particle system.

Fortunately there always exist unitaries which transform BT systems into
cluster-separable systems without affecting the S-matrix. Such transformations
are called \textit{scattering equivalences}. The phenomenon behind this
observation is vaguely reminiscent of the insensitivity of the S-matrix in QFT
against local changes in the interpolating field-coordinatizations which in
field theoretic terminology means changing the pointlike field by passing to
another (composite) field in the same Borchers class, or in the setting of
AQFT by picking another operator from a local operator algebra\footnote{The
class of fields which interpolate the same S-matrix is much larger (it
comprises all ''allmost local''\ fields \cite{Haag}) but in a local theory
there is no demand for such generality.}. The notion of scattering
equivalences is conveniently described in terms of a subalgebra of
\textit{asymptotically constant} operators $C$ defined by
\begin{equation}
lim_{t\rightarrow\pm\infty}C^{\#}e^{itH_{0}}\psi=0
\end{equation}
where $C^{\#}$ stands for both $C$ and $C^{\ast}.$ These operators which
vanish on dissipating free wave packets in configuration space form a
*-algebra which extends naturally to a C$^{\ast}$-algebra $\mathcal{C}.$ A
scattering equivalence is a unitary member $V$ of $\mathcal{C}$ which is
asymptotically equal to the identity%
\begin{equation}
lim_{t\rightarrow\pm\infty}(V^{\#}-1)e^{itH_{0}}\psi=0
\end{equation}
The relation to scattering theory comes about through the change of the M\o
ller operators according to $\Omega_{\pm}(VHV^{\ast},VH_{0})=V\Omega_{\pm
}(H,H_{0})$ which leaves the S-matrix unchanged. Scattering equivalences do
however change the interacting representations of the Poincar\'{e} group
$U(\Lambda,a)\rightarrow VU(\Lambda,a)V^{\ast}$

It has been shown by Sokolov \cite{Soko} that these scattering equivalences
can be utilized for achieving cluster separability of the (interacting)
representation of the Lorentz group while maintaining the S-matrix. For
example if we were to take the BT representation for the full 3-particle mass
operator \ (\ref{additive}) then there exists a unitary $B$ which does the
following%
\begin{equation}
H_{cl}=BH_{BT}B^{\ast}%
\end{equation}
where $H_{cl}$ denotes the Hamiltonian associated with the clustering
representation. In general these B's have a complicated (non-algebraic)
functional analytic dependence on the interaction data in $H_{BT}$ since
scattering theory enters in their calculation in an essential way. The
simplest case for an explicit construction is the above mentioned case of
3-particles with one spectator. which also serves as simple demonstration of
the failure of clustering in the BT setting \cite{Wayne}.

For a consistent formulation including bound states involving more than two
interacting particles, one needs the theory of rearrangement collisions which
uses in addition to the Hilbert space of the interacting system $\mathcal{H}$
an auxiliary Hilbert space $\mathcal{H}_{f}$ describing the free moving stable
fragments (elementary and bound particles). Since the Hamiltonian of the free
dissipating fragments and the actual Heisenberg Hamiltonian operate in
different Hilbert spaces we need an isometric map $\Phi$ which relates the two
in such a way that the M\o ller operators
\begin{equation}
s-lim_{t\rightarrow\pm}e^{itH}\Phi e^{-itH_{f}}=\Omega_{\pm}(H,\Phi,H_{f})
\end{equation}
are now isometries with the same ranges between the two spaces and $S=$
$\Omega_{+}(H,\Phi,H_{f})^{\ast}\Omega_{-}(H,\Phi,H_{f})$ is a unitary
operator in $\mathcal{H}_{f}.$ In case of absence of bound states (or for two
particles after projecting onto the subspace of scattering states) one may
choose $\Phi=1$ and the formalism reduces to that in one Hilbert space
$\mathcal{H=H}_{f}$. Although there is a lot of freedom in the choice of
$\Phi,$ some knowledge about bound state wave function is necessary.

The proof that a clustering Poincar\'{e} invariant and unitary S-matix exists
(for a given 2-particle interaction) is inductive \cite{C-P}. The induction
starts with two particles as above, where the BT additivity and the cluster
property hold simultaneously. As stated above, the cluster property does not
hold for 3 particles with one particle in a spectator position. The induction
assumption is

\begin{itemize}
\item  For each proper subsystem $\mathcal{C}_{i}$ of an N-body system there
is a representation $U_{cl,\mathcal{C}_{i}}(\Lambda,a)$ which clusters and is
scattering equivalent to a BT representation $U_{cl,\mathcal{C}_{i}}%
(\Lambda,a)=B_{\mathcal{C}_{i}}U_{BT,\mathcal{C}_{i}}(\Lambda,a)B_{\mathcal{C}%
_{i}}^{\ast}$ (remember the BT system has a mass operator $M_{BT,\mathcal{C}%
_{i}}$ which commutes with the $\vec{X}_{0},\vec{J}_{0},\vec{P}$ associated
with $\mathcal{C}_{i})$

\item  Let $\mathcal{P=}\cup_{i}\mathcal{C}_{i}$ be a partitioning of the
N-body system into proper subsystems and let $U_{\mathcal{P}}(\Lambda,a)=%
%TCIMACRO{\dprod \limits_{i}}%
%BeginExpansion
{\displaystyle\prod\limits_{i}}
%EndExpansion
$ $U_{cl,\mathcal{C}_{i}}(\Lambda,a)$ a representation of the Poincar\'{e}
group associated with the cluster partition $\mathcal{P}.$
\end{itemize}

\begin{theorem}
(Coester-Polyzou, \cite{C-P}\cite{Wayne2}) $U_{\mathcal{P}}(\Lambda,a)$ is
scattering equivalent to a BT system $U_{BT,\mathcal{P}}(\Lambda,a)$
corresponding to a mass (invariant energy) operator $M_{BT,\mathcal{P}}$
\end{theorem}

The last step consists in writing down the BT mass operator for the full
N-body system in terms of an additive formula which is the N-particle analog
of (\ref{additive}) and then constructing another scattering equivalence $B$
which transforms this N-body BT system into a system which clusters for
\textit{any} partitioning
\begin{equation}
s-lim_{\left\|  a_{i}\right\|  \rightarrow\infty}(U_{cl}(\Lambda,a)-\prod
_{i}U_{cl,\mathcal{C}_{i}}(\Lambda,a))e^{i\sum_{i}P_{\mathcal{C}_{i}}a_{i}}=0
\end{equation}
where the last factor denotes the translation which shifts the clusters of the
chosen clustering to infinity ($P_{\mathcal{C}_{i}}$ is the momentum of the
i$^{th}$ cluster). The induction starts at N=2 where all the properties are
obviously fulfilled.

The direct particle interaction setting of Coester and Polyzou incorporates
all those aspects of particle physics which can be formulated without local
fields. Micro-causality, locality, the crossing property of the S-matrix do
not belong to these properties, whereas Poincar\'{e}-invariance and the
cluster separability of the scattering operator are naturally incorporated. In
addition there are a number of properties which are compatible but cannot be
derived as structural consequences in this setting as: existence of
antiparticles, TCP symmetry, spin and statistics\footnote{According to Kuckert
\cite{Kuckert} a spin statistics connection can be derived in the setting of
nonrelativistic QM which possibly lacks a second quantization formulation.
This suggsts the interesting question whether these kind of arguments can be
carried over to the Coester-Polyzou direct relativistic particle interaction
setting.}. The direct particle interaction theory has been included in this
anthology of non-local particle theories, because the assumptions on which it
is founded define a catalogue of properties which are indispensable in any
theory of relativistic particles, local or non-local.

It is worthwhile to mention that Dirac's hole model of particles/antiparticles
interacting with a quantized electromagnetic field (which led to many low
order correct perturbative results, see Heitler's book) despite its undeniable
successes is not a correct theory of interacting particles and fields; this is
the reason why it was not possible to do renormalization theory in the setting
of hole theory. The reason is that (apart from some d=1+1 integrable models)
the processes of filling the ''sea''\ and switching on the interaction do not
commute, although in graphs without loops (vacuum polarizations) are not
affected. The C-P direct particle interaction theory is the only known
consistent relativistic particle framework which avoids the use of quantum
fields and maintains consistency on the level of particles. It is non-local
because the only notion of localization which (as in nonrelativistic theory)
can be expressed in terms of localizing projectors (the Newton-Wigner
localization) is only asymptotically L-covariant and quasi-local. But this is
enough to obtain a L-invariant and clustering S-matrix. In fact Wigner
particles only enter QFT through the LSZ asymptotics; the limitation of
localization resulting from the non-existence of finitely localized covariant
projection operators does not cause harm to covariance of scattering theory.

The description of finite propagation, as distinguished from asymptotic
scattering, requires a different localization which is inherent in QFT, but
has also a well-defined intrinsic meaning on particle states (related to the
modular localization in section 4). It does not play a role in the
Coester-Polyzou setting which is basically a scattering theory of relativistic
particles and not a description of propagation over finite spacetime
separations. An antagonism as pronounced as expressed in the title
``\ Reeh-Schlieder defeats Newton-Wigner''\ \cite{Halvor} does not really
exist. Instead of emphasizing the lack of covariance and causal localizability
of localization concepts build on the existence of projectors onto subspaces
\cite{Mala} and in particular of the N-W particle localization, it would have
been more helpful for particle physics to emphasize that they are
asymptotically covariant and causal and that they are indispensable for the
derivation of scattering theory were only the asymptotic localization
properties are relevant. The statement that their use in propagation over
finite times creates contradictions to Fermi's two-atom Gedankenexperiment
(designed to illustrate the validity of the maximal velocity principle in the
relativistic quantum realm \cite{Heger}) is quite irrelevant; the causal
propagation aspects are to be described by the modular localization which
leads to projectionless expectation values \cite{B-Y}\cite{Yng}.

It is also important to note that scattering theory and asymptotic
completeness lead to the description of the Hilbert space of the theory in
terms of a Fock space of multi-particle states. Whereas field
coordinatizations are (as coordinates in geometry) highly arbitrary, particles
belong to the intrinsic and unique content of particles physics.

QFTs as those one encounters in generic curved spacetimes, lack this powerful
particle aspect; they even possess no vacuum state. A similar loss of particle
structure is encountered if one passes from a ground state to a thermal
situation by placing the system into a heat bath \cite{Br-Bu}; instead of
Wigner particles, one confronts an ensemble of dissipating quanta which do not
comply with scattering theory nor relate to a Fock space structure. This
situation also prevails if the thermal aspect is generated by localization via
a causal horizon as in the Hawking-Unruh effect. There is simply no LSZ
scattering behavior and no Fock space structure associated with Unruh
excitations\footnote{The quantization of an interaction-free theory on Rindler
spacetime of course has the Fock space structure of the free quanta, but
without the validity of the LSZ scattering theory in a Rindler world this does
not extend to the interacting case.}. The hallmark of an Hawking-Unruh
situation, as compared to a generic heat bath caused thermality, is that there
exists a ``territorial''\ extension \cite{S-W2}\cite{G-L} beyond ``horizons''
which is accompanied by an extension of the type III$_{1}$thermal von Neumann
algebra with a KMS state at the Hawking-Unruh temperature to a standard type I
ground state algebra with a Wigner particle structure and an associated
scattering theory which permits to describe the Hilbert space of the theory in
terms of a Fock space of incoming particles; this extension is only possible
at that particular temperature. Note that our use of the notion of (Wigner)
particles, as those objects in terms of which all field states above the
ground state can be asymptotically resolved, is more restrictive that in
\cite{Wald}\cite{Clifton}.

\section{String-localized fields, stability under perturbations}

The simplest illustration of string-localization results from the association
of a localized quantum field to the famous class of Wigner's zero mass
representation of the Poincar\'{e} group for which the stability group (the
''little group'') of a lightlike vector has a faithful representation. Instead
of one helicity, as in the conformally invariant cases of photons and massless
neutrinos, this representation contains a tower of all integer- or
halfinteger-valued helicity degrees of freedom of both signs and we will
therefore refer to this class of representations (which depends on one
continuous parameter $\kappa,$ a kind of euclidean mass) as the helicity tower
or ``infinite'' spin representation \cite{M-S-Y}.

It had been known for some time that there can be no pointlike covariant field
which applied once to the vacuum leads to a one-field subspace containing such
a representation \cite{Y}. In more recent times, the application of the
spatial version of modular theory applied to positive energy representation
via forming intersections of wedge-localized subspaces in positive energy
Wigner representations has led to a theorem that they are always localizable
in (arbitrary thin) spacelike cones \cite{B-G-L}. Since the cores of such
cones are semiinfinite linear spacelike strings (in analogy to points being
the cores of compact double cone), it is natural to look for a
string-localized field as being the best (tightest localized) field theoretic
description of Wigner's zero mass infinite helicity tower representations.
This object indeed exists and in the following we will describe this
construction in some detail.

The irreducible \textit{zero mass, infinite helicity-tower representations} of
the orthochronous proper Poincar\'{e} group $\mathcal{P}_{+}^{\uparrow}$ are
induced from unitary irreducible representations of its stabilizer subgroup of
a fixed light-like vector (the ``little group'') . The stabilizer group in
this case is isomorphic to the two-dimensional \ Euclidean group $E(2)$,
consisting of rotations $R_{\vartheta}$ by an angle $\vartheta\in\mathbb{R}$
mod $2\pi$ and translations by $c\in\mathbb{R}^{2}.$ Let $\varphi_{1}%
\cdot\varphi_{2}=\int\delta(\left|  k\right|  ^{2}-\kappa^{2})\overline
{\varphi_{1}(k)}\varphi_{2}(k)~($where the bar denotes complex conjugation) be
the scalar product on the Hilbert space $H_{\kappa}$ of functions on the
plane, restricted to the circle of radius $\kappa$. An irreducible unitary
action of $E(2)$ on $H_{\kappa}$, with the Pauli-Lubanski parameter $\kappa,$
which is the Casimir invariant labelling nonequivalent representations, is
given by the formula
\begin{equation}
\left(  D_{\kappa}(R_{\vartheta},c)\varphi\right)  (k)=e^{ick}\varphi
(R_{\vartheta}^{-1}k)
\end{equation}
where $(R_{\vartheta},c)\in E(2)$ are the Euclidean rotation and translation
and $\kappa$ is a kind of Euclidean mass. The representation can be linearized
by Fourier transformation with respect to k.

Let $\psi(p)$ be an $H_{\kappa}$-valued wave function of $p\in\mathbb{R}^{4},$
square integrable with respect to the Lorentz invariant measure $d\mu
(p)=\theta(p^{0})\delta(p^{2})$ on the mantle $\partial V^{+}$ of the forward
light cone $V^{+}$. The unitary Wigner transformation law for such a wave
function reads
\begin{equation}
U(a,\Lambda)\psi(p,k)=e^{ipa}D_{\kappa}(R(\Lambda,p))\psi(\Lambda^{-1}p)
\end{equation}
where $R(\Lambda,p)=B_{p}^{-1}\Lambda B_{\Lambda^{-1}p}\in E(2)$ denotes the
Wigner ``rotation''\ with $B_{p}$ an appropriately chosen family of Lorentz
transformations that transform the standard vector $p=(1,0,0,1)$ to
$p\in\partial V^{+}.$ The Fourier series decomposition of $k\in S_{\kappa}%
^{1}$ leads to the discrete integer-valued helicities. Different from the zero
mass finite spin representations, the occurrence of the opposite helicity in
the infinite helicity tower makes a doubling of the representation (in order
to achieve TCP invariance) unnecessary. A valuable hint as to how to
investigate the localization aspects of this representation comes from placing
it into the setting of tensor product representations of the form
\begin{equation}
U^{0}(\Lambda,a)\otimes U^{1}(\Lambda) \label{tensor}%
\end{equation}
acting on the tensor product $H^{0}\otimes H^{1}$ of the representation space
of a spinless massive representation of the Poincar\'{e} group tensored with a
unitary representation of the homogeneous Lorentz group. The latter is
precisely the setting for the Bros-Morschella localization on 3-dimensional de
Sitter space which can be made explicit with the help of the H\"{o}%
rmander-Fourier transformation \cite{B-M}. The De Sitter space localization
arising from the second factor implies a spacelike directional localization in
the d=1+3 Minkowski spacetime which together with the pointlike localization
coming from the first factor amounts to a localization along semiinfinite
spacelike strings. The analytic mathematical details are somewhat more
demanding as for massive representation and the reader is referred to the
literature \cite{M-S-Y}.

The associated string-localized field operators are defined on the Fock-space
over the irreducible representation space and turn out to be of the form
\begin{align}
&  \Phi^{\alpha}(x,e)=\int_{\partial V^{+}}d\mu(p)\left\{  e^{ipx}u^{\alpha
}(p,e)\cdot a(p)+e^{-ipx}\overline{u^{\alpha}(p,e)}\cdot a^{\ast}(p)\right\}
\\
&  D_{\kappa}(R(\Lambda,p))u^{\alpha}(\Lambda^{-1}p,e)=u^{\alpha}(p,\Lambda
e)\nonumber\\
&  u^{\alpha}(p,e)\equiv e^{-i\pi\alpha/2}\int d^{2}ze^{ikz}\left(  B_{p}%
\xi(z)\cdot e\right)  ^{\alpha}~\nonumber\\
&  with\,\,\,\xi(z)=(\frac{1}{2}\left(  \left|  z\right|  ^{2}+1\right)
,z_{1},-z_{2},\frac{1}{2}\left(  \left|  z\right|  ^{2}-1\right)  )\nonumber
\end{align}
where the intertwiner $u^{\alpha},$ which depend on a complex parameter
$\alpha$\footnote{Although the Bros-Moschella setting fixes the parameter
$\alpha$ to $\operatorname{Re}\alpha=-1,$ the intertwiners which convert the
Euclidean degrees of freedom into internal fluctuating string degrees of
freedom exist (in the distribution theoretical) sense for all $\alpha;$ in
fact the continuous parameter $\alpha$ characterizes the linear part of the
string Borchers class.} (and on $p$ via the boost $B_{p})$ are determined by
the intertwining property in the second line and certain analyticity
requirements (with a complex parameter $\alpha$ to be explained). The dot
between the pre-factors $u^{\alpha}(p,e)$ and the creation and annihilation
operators $a^{\ast}(p),a(p)$ (that depend also on $k\in\mathbb{R}^{2}$,
suppressed by the notation) stands for integration over $k$ with respect to
the measure $\delta(\left|  k\right|  ^{2}-\kappa^{2})d^{2}k.$ The
$k$-dependence of $u^{\alpha}$, $a$ and $a^{\ast}$ has thus been transferred
to the dependence of the field $\Phi(x,e)$ on the space-like direction $e$.
The field is an operator-valued distribution in $x$ and $e$. In more physical
terms $\Phi(x,e)$ is a quantum field which fluctuated in the vacuum state in
4-dimensional Minkowski space as well in 3-dimensional de Sitter space (i.e.
there is a quantum localization in both spaces \cite{B-M}). It has the
following properties that justify the terminology ``string-localized'':

\begin{itemize}
\item \bigskip If $x+\mathbb{R}^{+}e$ and $x^{\prime}+\mathbb{R}^{+}e^{\prime
}$ are space-like separated then
\begin{equation}
\left[  \Phi^{\alpha}(x,e),\Phi^{\alpha^{\prime}}(x^{\prime},e^{\prime
})\right]  =0 \label{loc}%
\end{equation}
while the commutator is nonzero for some $e,e^{\prime}$ if the endpoints only
are space-like separated

\item  The transformation law of the field is consistent with this
localization:
\begin{equation}
U(a,\Lambda)\Phi^{\alpha}(x,e)U(a,\Lambda)^{-1}=\Phi^{\alpha}(\Lambda
x+a,\Lambda e) \label{cov}%
\end{equation}

\item  After smearing with compactly supported test functions in $x$ and $e$
(in certain $\alpha$-ranges a smearing in $e$ is unnecessary), the field
operators generate a dense set in Fock space when applied to the vacuum vector
$|0\rangle$. This is the appropriately string-adapted version of the
Reeh-Schlieder \cite{S-W} property which plays a crucial role in the
mathematical physics literature on algebraic QFT.
\end{itemize}

The second statement is (as in the standard finite spin case) a result of the
above intertwining properties of $u^{a}(p,e)$. This intertwiner function must
have a certain complexity, since according to the first above property $\Phi$
must amalgamate Minkowski and De Sitter spacetime in such a way that the
commutation properties are not simply those of a tensor product between
two-point function in both spaces\footnote{In particular it cannot be reduced
to standard type Jordan-Pauli type of zero mass commutators.} (which would
have too strong commutation properties). For an explicit representation of
$u^{a}(p,e)$ in terms of the H\"{o}rmander-Fourier transformation on de Sitter
space we refer to \cite{M-S-Y}.

Although it does not seem to be possible to express the string-localization
intertwiner $u^{\alpha}$ and the resulting c-number commutator in terms of
known functions, its existence and analytic properties can be proven from the
representation. It turns out that the continuous parameter $\alpha$ (apart
from the integer nonnegative values 0,1,..) can be chosen at will and the
different relatively local string fields in the same Fock space correspond to
the linear part of the local equivalence (Borchers) class. They are the
analogs of the discrete family of fields associated to the standard (m,s)
Wigner representation by taking $u(p,s)$ and $v(p,s)$ intertwiners of
increasing length $\ l\geq2s+1$\cite{Wein}. Presently the important question
of whether the operator algebra generated by these string-localized fields
contains compactly localizable subalgebras (possibly generated by pointlike
localized composite fields) is under investigation \cite{Project}.

Although the same construction with $\alpha$-dependent intertwiners
$u^{\alpha}(p,e)$ applies as well to the massive $(m,s)$ representation and
leads to semiinfinite string-localized fields, their application to the vacuum
generates the same one-particle Wigner representation spaces as the
$e$-independent standard textbook intertwiners $u(p)$ which belong to
pointlike fields. They could however play a role in enlarging the
possibilities for interactions. Whereas the case of interacting pointlike
localized fields \cite{Wein} has been studied since the very beginning of
field theory (and led to the result of renormalized perturbation theory in
which the pointlike localization is maintained in every order), little is
known about the interactions of string-localized fields.

There are indications that if one implements interactions with polynomials of
string fields $A(x,e),$ the perturbation will maintain string stability in the
sense that the resulting fields will not have a localization which spreads
beyond the original string. This makes the string-like localization an
excellent candidate for a type non-local theory which still maintains an
extended form of causality. For zero mass finite spin representation the
possibilities of pointlike free fields are more restricted than in the massive
case. For example the Wigner representation for photons admit a description in
terms in terms of pointlike field strength but not in terms of a pointlike
vector potential. The only possibility for a physical (i.e. one without
unphysical gauge-dependent degrees of freedom) vector potential is a string
localized vector potential \cite{Project}. This suggests to use string
interactions as an interesting alternative to gauge interactions.

Another type of string-localized field arises from the massive Wigner
representation in d=1+2 for (abelian) generic real values of the spin
(different from (half)integer). Such generic ``anyonic'' values activate the
rich covering structure of the 3-dimensional Lorentz group. The covering
structure beyond the double covering requires to define real localization
subspaces more carefully by attaching a spacelike direction $e$ such that
$W+e\subset W$ \cite{Mund3}$.$ The directions $e$ are again points in a De
Sitter space, but since this space is now two-dimensional this de Sitter space
has an infinite covering. In this way the pairs $\tilde{W}\equiv(W,e)$ and
finally also the sharpened pointlike data $(x,e)$ can be used to model a
substitute for covering space of spacelike wedges respectively of spacelike
strings which matches the Bargmann covering structure of the Lorentz
group\footnote{Another way to associate a covering space with the Minkowski
spacetime is to compactify Minkowski space and then take its universal
covering. The symmetry group of the covering is the the covering of the
conformal extension of the Poincar\'{e} group. the presence of covering groups
beyond matrix groups implies the presence of interactions (see third section).
The timelike covering of the compactification does not lead to strings but
rather to a timelike braid group structure.}. The geometric role of the
spacelike directions is to have a substrate on which the center of the
covering can act nontrivial. In this way the string nature of the resulting
objects is already preempted by the covering structure of the modular wedge
formalism which requires the definition of a reference wedge from which the
``winding number''\ is counted. One easily shows that unlike the (half)integer
spin case there is now a complex phase factor between the symplectic
complement of $H_{r}(\tilde{W}),\,\tilde{W}:=(W,e)$ and its geometric
complement i.e. $H_{r}(\tilde{W})^{\prime}=Phase\cdot H_{r}(\tilde{W}^{\prime
}).$ Since the symplectic complement is the one-particle projection of the von
Neumann commutant and the geometric complement is given by the square root of
the $2\pi$ spin rotation, the phase is obviously related to the
spin-statistics phase of abelian braid group commutation relations (anyons).
The presence of string localization can be analytically confirmed by showing
the complex phase requires the triviality of compactly localized subspaces.
The smallest intersections of wedges which are consistent with $\mathcal{C}%
+e\subset\mathcal{C}$ are spacelike cones with arbitrary opening angles.

This kind of string (referring to the ``would be''\ string-like generators of
the spacelike cone localized field algebras) is very different from the
previous one since it owes its string-like nature not to the presence of an
internal structure of the little group with internal string degrees of
freedom. Fields transforming according to higher (than two-fold) coverings of
$SO(2,1)$ carry no more degrees of freedom than Bosons or Fermions. Whereas
the anyonic string resembles what one expects from Mandelstam strings (the
oldest use of gauge invariant strings in QED is due to Jordan \cite{Jordan})
i.e. strings in gauge theories\footnote{There are also the semiinfinite
Buchholz-Fredenhagen strings \cite{B-F} of local quantum physics coming from
the classification of admissible superselection charges in the presence of a
mass gap. The authors believe that they are the model independent and rigorous
version of strings in massive gauge theories with confinement.}, the strings
with the internal degrees of freedom in the form of an \textit{infinite
helicity tower} are in some sense more like the objects of string theory (see
however the caveats below). The anyonic strings are for various reasons (e.g.
potential applications in solid state physics) of more immediate physical
interests than the previous zero mass infinite spin strings, but their highly
nontrivial field theory (there are no on-shell free fields) will not be
considered here.

String-localized objects are radically different from the properties of
quantized Nambu-Goto strings through which Polyakov found the string
interpretation of the dual model in terms of functional integrals. There exist
two different quantizations of the N-G string. The more intrinsic approach is
due to Pohlmeyer \cite{Po}\textit{ }and consists in extracting a complete set
of classical invariants (called Pohlmeyer charges) which together with the
generators of the Poincar\'{e} group form a closed Poisson algebra. This task
has been almost completed, and in this way the system was identified with an
integrable system in the classical sense. The quantization of this
algebra\footnote{It is believed that the correct quantum theory for integrable
systems is obtained by quantizing its invariant charges.}, i.e. the search for
an algebra of operators which have commutation relations which mimic the
Poisson structure as much as possible, is still an unfinished problem. The
Pohlmeyer strings are not string-localized, but their algebra of invariants
share with the string-localized fields that they exist as Poincar\'{e}
covariant theories in all spacetime dimensions $d\geq3$ i.e. there is no
mysterious distinction (resulting from the non-intrinsic canonical
quantization) of $d=10,26.$

The other approach is that which led to string theory proper and consists in
quantizing before eliminating the constraints and afterwards taking care of
the constraints in the spirit of the BRST cohomological approach. The
localization aspect was investigated \cite{Ma}\cite{Di} and it was found that
their quantum localization defined in terms of commutators is pointlike,
unlike its classical appearance. There exist also recent arguments that the
S-matrix of string field theory can be associated with a (possible infinite)
family of local fields \cite{Erler}.

Since the quantization of a classical Lagrangian in its quantization should
not lead to two different QT, this raises the question whether there is a
right one and which is it. From the viewpoint of quantization of integrable
systems it is Pohlmeyer's quantization; however the string theoretic reading
of the dual model requires the use of the canonical quantization and the BRST formalism.

\section{Wedge-localized PFGs}

Non-local auxiliary operators play a pivotal role in nonperturbative (and
non-Lagrangian) constructions of QFT models. Their purpose is to generate
operator algebras which are localized in extended causally closed regions. The
best studied case is that of wedge-localized generators without vacuum
polarization, which lead among other things to the d=1+1 Zamolodchikov-Faddeev
algebras of factorizing models.

Let us briefly look at this interesting new aspects by explaining the meaning
of vacuum-\textbf{p}olarization-\textbf{f}ree \textbf{g}enerators shortly
called PFGs. Let $\mathcal{O}$ be a causally closed, simply connected region
in Minkowski spacetime, typically (in order of decreasing size) a wedge
$\mathcal{W}$, a spacelike cone $\mathcal{C}$ or a compact double cone
$\mathcal{D}$. A $\mathcal{O}$-associated PFG $F$ is a not necessarily bounded
operator affiliated with the operator algebra $\mathcal{A}$($\mathcal{O}$)
(denoted by $F\eta\mathcal{A}(\mathcal{O}))$ such that the vector obtained by
applying $F$ once to the vacuum $F\Omega$ contains only a one-particle
component i.e. has no vacuum multiparticle polarization parts. The operator
algebras of a free field theory clearly possess PFGs for every spacetime
region. However wedge-localized PFGs even exist in interacting theories (to
the extend that have isolated one-particle states). This is a general
consequence of modular theory. But in most cases these PFG are not useful
because their translates do not admit Fourier transforms and their extremely
bad domain properties prevent their successive applications as in Wightman
field theory.

The useful PFG are the so-called ''tempered PFG''.\ They generally do not
exist in the presence of interactions, even if we allow the region to be as
large as a wedge. In fact if a PFG $F$ exists for a subwedge region (like a
spacelike cone $C$ or a double cone $\mathcal{D}$), then it follows that the
$F$ is in fact a smeared free field (with support of the test function in the
localization region). The analytic part of the proof is very similar to the
proof of the Jost-Schroer theorem \cite{S-W}, the slight complication due to
the fact that the operator $F$ has a prescribed localization property but no
simple (tensorial) covariance behavior is easy to account for.

So the non-existence of subwedge PFG is the precise intrinsic local
characterization for the presence of interactions which turns out to be
equivalent (under the assumption of validity of the crossing property of
formfactors) to $S_{scat}=1.$ Modular theory establishes the existence of
wedge-localized PFGs, but PFGs in the stronger sense of temperateness only
exist in case of interactions without real (on-shell) creation and
annihilation in scattering processes. It turns out that a purely elastic
$S_{scat}$ are only possible in d=1+1 and it is believed that this possibility
is exhausted by the family of factorizing models. Schematically, limiting the
notation to the simplest case of a scalar scattering matrix in a factorizing
theory involving one particle only (e.g. the Sinh-Gordon model) we have for
the wedge localized PFG $A(x)$ the following on-shell representation
\cite{schroer}
\begin{align}
A(x)  &  =\int(Z(\theta)e^{ip(\theta)x(\chi)}+h.c.)d\theta\\
p(\theta)  &  =m(ch\theta,sh\theta),~x(\chi)=r(sh\theta,ch\theta)\nonumber
\end{align}
where the second line defines the mass shell rapidity parametrization as well
that of the right wedge. The $Z^{\#}s$ are not the standard
creation/annihilation operators, but they rather fulfill the
Zamolodchikov-Faddeev algebra relations:
\begin{align}
Z(\theta)Z^{\ast}(\theta^{\prime})  &  =S(\theta-\theta^{\prime})Z^{\ast
}(\theta^{\prime})Z(\theta)+\delta(\theta-\theta^{\prime})2\\
Z(\theta)Z(\theta^{\prime})  &  =S(\theta^{\prime}-\theta)Z(\theta^{\prime
})Z(\theta)\nonumber
\end{align}
Therefore $A(x)$ can not be a local field (except $S(\theta)=\pm1$). Some
application of modular theory of operator algebras for the wedge region shows
that
\begin{align}
A(f)  &  =\int A(x)f(x)d^{2}x,~\sup pf\subset W\\
\mathcal{A}(W)  &  =alg\left\{  A(f),~\sup pf\subset W\right\} \nonumber
\end{align}
generate the wedge algebra.

In contradistinction to pointlike localized fields the sharpening of the
support of $f$ to a subregion inside $W$ does not help since the localization
region of resulting operator does not follow the classical picture of the
support of $f$. The only way to restrict the localization is to define
algebras with the more restricted localization by forming the intersections of
wedge algebras which contain $D$ i.e.
\begin{equation}
\mathcal{A}(D)=\cap_{W\supset D}\mathcal{A}(W)
\end{equation}
which for the case of d=1+1 at hand can be shown to reduce to
\begin{align}
\mathcal{A}(D)  &  =\mathcal{A}(W)\cap\mathcal{A}(W_{a})^{\prime}\\
D  &  =W\cap W_{a}^{\prime}\nonumber
\end{align}
the condensed notation requires some explanation. $W_{a}$ is the result of
applying a right spacelike translation by $a$ to $W$ and the upper dash on
regions denotes their spacelike complement which in the case of a wedge is
just the opposite left wedge. The same dash but on operator algebras denotes
the von Neumann commutant in the ambient Hilbert space. The computation of
intersections of von Neumann algebras is a complicated problem for which no
general methods exist. However if the generators of wedge algebras have a
fairly simple structure as in our case, one can solve the characterizing
relations for the desired $\mathcal{A}(D)$ affiliated operators $A$ as those
which obey the relation
\begin{equation}
A\subset A(W)\ s.t.\left[  A,U(a)A(f)U(a)^{-1}\right]  =0 \label{commute}%
\end{equation}
In the spirit of the old LSZ formalism one can then make an Ansatz in form of
a power series in $Z(\theta)$ and $Z^{\ast}(\theta)\equiv Z(\theta-i\pi)$
(corresponding to the power series in the incoming free field in LSZ theory)
\begin{equation}
A=\sum\frac{1}{n!}\int_{C}...\int_{C}a_{n}(\theta_{1},...\theta_{n}%
):Z(\theta_{1})...Z(\theta_{n})
\end{equation}
Each integration path $C$ extends over the upper and lower part of the rim of
the $(0,-i\pi)$ strip. The strip-analyticity of the coefficient functions
$a_{n}$ expresses the wedge-localization of $A.$ It is easy to see that these
coefficients are identical to the vacuum polarization form factors of $A$%
\begin{equation}
\left\langle \Omega\left|  A\right|  p_{n},..p_{1}\right\rangle ^{in}%
=a_{n}(\theta_{1},...\theta_{n})
\end{equation}
whereas the crossing of some of the particles into the left hand bra state
leads to the connected part of the formfactors
\begin{equation}
^{out}\left\langle p_{1},..p_{l}\left|  A\right|  p_{n},..p_{l+1}\right\rangle
_{conn}^{in}=a_{n}(\theta_{1}+i\pi,...\theta_{l}+i\pi,\theta_{l+1}%
,..\theta_{n})
\end{equation}
The algebraic identity (\ref{commute}) involving the in $Z$ linear generator
$U(a)A(f)U(a)^{-1}$ relates the $a_{n}$ coefficients whose n differs by 2.
This is nothing else than the famous ''kinematical pole condition'' first
introduced as one of the construction recipes by Smirnov \cite{Smir}. The
solutions together with the Payley-Wiener meromorphic characterization of the
size of $D$ defines a space of formfactors or $A^{\prime}s$ in the sense of
bilinear forms.

If one wants $A$ to have simple covariance properties one should think of a
basis of pointlike fields (the generalization of Wick monomials to the realm
of interaction) which in this context does not create short distance problems
since formfactors by definition do not contain short distance fluctuations.
Hence if we would know that the intersected algebras $\mathcal{A}(D)$ are
nontrivial then the present approach serves as a calculational tool for the
formfactors of these operators. Direct calculations of correlation functions
with the aim of showing the existence via a GNS kind of reconstruction theorem
failed because one has not been able to control the sum over intermediate
particle states which formally links correlation functions of products of
operators with formfactors of these operators.

In this situation it is interesting to note that there has been a recent
proposal by Buchholz and Lechner to use powerful modular methods of operator
algebras (''modular nuclearity'') instead if handling infinite intermediate
state sums using formfactors. If this modular nuclearity property can be
checked for the wedge algebras than the nontriviality of the intersection is
guarantied and one would have succeeded to proof the existence of a large
class of nontrivial models whose fate would otherwise fall under the spell of
short distance properties.

The idea in this form depends on the existence of nontrivial wedge-localized
PFG (only possible in absence of inelastic scattering processes) and can not
be generalized to non factorizing models or to higher dimensions. Since the
presentation of some highly speculative ideas how to get around these problems
would lead to far away from the main subject of this paper we refer to
forthcoming work.

On the other hand Brunetti Guido and Longo \cite{B-G-L} constructed the family
of wedge-localized real subspaces of the Wigner representation spaces by
combining purely group theoretical ideas with aspects of modular theory.
Although the subwedge (spacelike cones, double cones i.e. the natural simply
connects causally complete regions obtained by intersecting wedges) localized
real subspaces have no geometric modular characterization, they can easily be
constructed by intersecting wedge-localized spaces. The work of these authors
culminated in a theorem that positive energy representations of the
Poincar\'{e} group always have nontrivial spacelike cone-localized subspaces
for arbitrary small opening angles of the spacelike cone and that the wedge
like localization can be re-obtained by additivity from spacelike cone
localized subspaces. In case of (half)integer spin representations one can
show that the compact spacelike cone intersections are nontrivial by
constructing a dense set of double cone localized wave functions. However
there were two class of representations which did not allow compact
localization (i.e. the double cone localization spaces were trivial), the
d=1+2 representations with ''anyonic''\ spin (s$\neq$ (half)integer) and the
famous Wigner family of zero mass infinite helicity representations. Both
cases lead to nontrivial spacelike cone localized subspaces and hence it was
suspected that these representations may be associated with string-localized
field theories. This turns out to be true; but only in the infinite spin case
the string-localized fields are genuine free fields, whereas in the anyonic
case the string localized fields applied to the vacuum are always accompanied
by vacuum polarization\footnote{Only wedge-localized algebras of anyon
theories support generators (tempered PFGs \cite{B-B-S}) without vacuum
polarization which can be viewed as being directly associated with wave
functions from the the corresponding wedge-localized Wigner representation
subspaces.} which throws them off mass shell.

The modular approach to the d=1+1 factorizing models is very similar in that
the principle structure is always associated with wedge-localized algebras and
the subwedge-localized objects are formed by intersections. The main
difference is that in all situations of positive energy Wigner representations
(with the exception of the case of d=1+2 anyons) there also exist
\textit{subwedge-localized PFGs} and their associated algebras are obtained in
a very simple functorial way by from a spatial modular theory in a functorial
way (the CCR or CAR functor). This had led me already in a very early stage of
my investigations \cite{S}\cite{S1} to view the whole constructive program
based on modular theory as an extension of the functorial relation between
spatially and algebraically localized objects to the realm of interactions.

In this section I have been using two different looking, but nevertheless
equivalent intrinsic definitions of the meaning of ''interacting''. On the one
hand one can take the absence of PFGs associated to subwedge-localized
algebras as the local definition of absence of interactions (and their
presence in the opposite situation). On the other hand there exists the global
definition in terms of the triviality of the S-matrix i.e. S=1. The modular
aspects of wedge-localized algebras permits to identify these two definitions.

\section{Non-local aspects of QFT in noncommutative Minkowski spacetime}

More than a decade ago Doplicher, Fredenhagen and Roberts \cite{D-F-R} started
to investigate the feasibility of a QFT on noncommutative Minkowski space.
Their original motivation was to study implications of certain uncertainty
relations which arose from limitations on measurements in small volumes if one
requires that no black hole horizons should be created by performing such
measurements alone i.e. without doing anything else. They then realized that
in case one ignores curvature these uncertainty relations can be saturated in
terms of a noncommutative Minkowski spacetime i.e. an algebra affiliated with
hermitian localization operators which fulfill (in a suitable normalization)%
\begin{align}
\left[  q_{\mu},q_{\nu}\right]   &  =i\lambda_{P}^{2}Q_{\mu\nu},\,\left[
q_{\lambda},Q_{\mu\nu}\right]  =0\label{Q}\\
Q_{\mu\nu}Q^{\mu\nu}  &  =0,\,(\frac{1}{16}Q_{\mu\nu}\widetilde{Q}^{\mu\nu
})^{2}=1\nonumber
\end{align}
where the matrix $Q$ is central-valued and transforms under L-transformation
like an antisymmetric tensor and $\widetilde{Q}$ the associated pseudo tensor.
The ''Planck length'' $\lambda_{P}$ is a parameter whose intended role is that
for large distances the spacetime is to return to Minkowski spacetime.

An irreducible representation of the $q^{\prime}s$ would require numerical
values $\sigma$ of $Q$ which would destroy L-invariance. The L-transformed
$q^{\prime}=\Lambda q$ leads to another irreducible representation with
another commutator matrix $\sigma^{\prime}$ which results from L-transforming
the antisymmetric tensor $\sigma_{\mu\nu}.$ The minimal $\ast$algebra on which
the Poincar\'{e} transformations can act as automorphisms is therefore a
direct integral over $\Lambda q$ with the Haar measure of the L-group. Since
the Lorentz group only enters the algebraic structure via its transitive
action on the $\sigma_{\mu\nu}$ matrices, the direct integral actually
involves a measure on $\Sigma$ which is induced by the Haar measure of the
L-group. It is comforting to know that this formal $^{\ast}$algebra on which
the Poincar\'{e} group acts as an automorphism can be elevated to a
mathematically respectable $C^{\ast}$-algebra whose regular representation is
unique (up to quasiequivalence) \cite{D-F-R}. The spectral values which the
central elements $Q$ take consists of all values $\sigma\in\Sigma$ with the
above restriction for the matrix entries; explicitly this set $\Sigma$ turns
out to be isomorphic to the 4-dimensional tangent space of a sphere extended
by a reflection $\Sigma\simeq TS^{2}\times\{-1,1\}.$

There exist Poincar\'{e} covariant representations of the algebra affiliated
with these commutation relations and the full spectrum $\Sigma$ (which simply
consists of the space of numerical matrices fulfilling the above algebraic
conditions (\ref{Q})) \cite{D-F-R} of the nontrivial center; this situation is
summarily referred a as the non-commutative Minkowski spacetime $M_{qu}$. The
representation of the translations (incorporation of the momentum space)
requires to tensor the Hilbert space $H$ which carries the $q$-representation
with its conjugate $H_{rep}=H\times\bar{H}$. Different from the positive
energy representations featuring in particle physics, the translational part
is not related to support properties in momentum space (in particular no
positive energy restriction).

What is meant by a field on such a noncommutative Minkowski spacetime can be
best explained in case of a free field\footnote{This restriction is not only
taken for pedagogical reasons; it is presently not quite clear whether the
standard implementation of interactions in terms of polynomials of
(non-commutative) field products is appropriate.}. Formally it is obtained by
substituting the spacetime variable by the above operator $q$%
\begin{align}
A(q)  &  =\int(e^{-ipq}a(p)+h.c.)\frac{d^{3}p}{2p_{0}}\\
H  &  =H_{rep}\otimes H_{Fock}\nonumber
\end{align}
where the second line denotes the Hilbert space of the free model and the
first factor is the previously explained covariant representation space. The
Poincar\'{e} transformations do not only act on the particle variables
(including the transformations of their ``non-commutative positions''), but
they also act on the central operators $Q$ and transform those parts of their
spectrum together the localizing wave functions. In this way we clearly keep
the Wigner particle picture as one of the pillars of any relativistic particle
theory. If we want to return to a Fock space description, we have to take the
(partial) expectation value in a state $\omega$ on the localization space
$M_{qu}$ which leads to expressions of the form \cite{D-F-R}\cite{Bahns} e.g.%
\begin{subequations}
\begin{align}
&  \omega\rightarrow\omega(A(q+x_{1})...A(q+x_{n}))=\label{lo}\\
&  =\int f_{\omega}(x_{1},x_{2},...x_{n};\xi_{1},\xi_{2},...\xi_{n})A(\xi
_{1})A(\xi_{2})....A(\xi_{n})d\xi_{1}d\xi_{2}...d\xi_{n}\nonumber
\end{align}
This leaves us with smeared non-local usual operator in Fock space (written in
terms of a non-local monomial of local free fields) where the non-locality
comes from collecting the product of $q$-exponentials in one exponential
only\footnote{The non-commutativity of the $q$-exponentials converts the
ordinary product of field operators into a ``twisted'' product. The twist
factors are center-valued and it is only their evaluation in a state $\omega$
on the $M_{qu}$ which leads to $\mathbb{C}$-valued weighted integrals over the
$\vec{Q}$-spectrum.} on which the state $\omega$ over the $q$-algebra may be
evaluated. If one starts e.g. with an $\omega$ corresponding to ``minimal
localization'' \cite{D-F-R}, the spectral values of $Q$ are restricted to a
compact submanifold $\Sigma_{1}$ which is left invariant by the rotation and
translation subgroups whereas a L-boost will change its position within
$\Sigma.$ Since L-invariant states on $M_{qu}$ do not exist, the field
theoretic expressions on $H_{Fock}$ are even before test-function smearing
already delocalized and non-covariant (in the sense that they are not fitting
the tensor/spinor calculus of Lagrangian quantization). The commutator of two
such fields evaluated on $\Sigma_{1}$ associated minimal states $\omega$
exhibits a Gaussian decrease in spacelike direction. This is faster than the
Yukawa decrease allowed by the Borchers-Pohlmeyer theorem mentioned in the
introduction, but one has to keep in mind that a free field theory on $M_{qu}$
is quite different from a Wightman theory on ordinary Minkowski spacetime $M$
so that the analyticity properties used in the derivation of the
Borchers-Pohlmeyer theorem are violated.

Although there is no problem with free noncommutative QFT, the interacting
situation is still in a very precarious state. The problem with interactions
is similar to those problems on which the old attempts (mentioned in the
introduction) failed.

The indispensable property which any particle physics theory hs to deliver,
whether P-covariant and non-local in the old sense or P-covariant and
non-local in the new sense of non-commutative Minkowski spacetime, is a
P-invariant S-matrix which describes scattering between Wigner multi-particle
in and out states.

Before one tries to formulate this requirement in terms of new interacting
fields on $M_{qu}$ in a LSZ setting of asymptotic convergence, one may try to
make mathematical experiments with a Gell-Man-Low like Ansatz for an S-matrix
or similar attempts with formulas close to the standard ones. The results in
this direction have been only partially successful. On the one hand it was
possible to overcome previous difficulties with unitarity and ultraviolet
divergencies \cite{B-D-F-P}. These perturbative calculations can even be
arranged in such a way that Feynman graphs supplemented with other rules
continue to be useful \cite{Piacitelli}. On the other hand it is presently not
known of how to obtain an S-matrix in the aforementioned intrinsic sense of
particle physics. It seems that the intuitive idea that noncommutative
Minkowski space asymptotically (for large distances) turns into standard
Minkowski spacetime does not work for on-shell quantities like the S-matrix
whose definition should be protected against short distances.

Note that in the C-P direct particle interaction theory the non-locality also
prevented the existence of covariant tensor objects corresponding to
Heisenberg fields\footnote{The theory is not capable to produce formfactors
i.e. particle matrix elements of covariant currents etc.}. But thanks to the
scattering equivalences this, did not spoil the existence of a clustering
unitary S-matrix. The D-F-R model is still best by the problem whether
interacting correlation functions in $\omega\times\omega_{vac}$ states will
fulfill cluster fall-off properties, which is a prerequisite for the existence
of a clustering S-matrix.

On the other hand one should perhaps not put too much faith in the particular
D-F-R model of fields on $M_{qu}$ but investigate more profoundly other
implementations of their physically well-motivated commutation relations.
Perhaps there is also a message to be drawn from the successful modular
localization setting. The representation of the Poincar\'{e}-group (including
reflections) on $H_{rep}\times H_{Wig}$ (where $H_{Wig}$ stands for the Wigner
irreducible particle representation) has also a wedge-associated Tomita
operator and therefore is subject to the setting of modular localization
(although with less restrictive properties since the representation in the
first factor is not of positive energy \cite{B-G-L}) which seems to allow for
a wider starting point than that defining the D-F-R model. \ 

\section{\bigskip Concluding remarks, outlook}

In this work we presented post renormalization attempts to formulate theories
in particle physics which either go beyond the frontiers of relativistic
locality (spacelike locality and timelike causality) but still maintaining
Wigner's particle concept, or proposals which aim at local theories but use
non-pointlike operators in intermediate steps as an essential constructive tool.

Since the word ''non-local'' is as non-revealing as a ''non-elephant'', one
needs to specify in what sense they non-local theories or non-local operators
deviate from locality.

This is most easy done in case of the wedge-localized objects (section 4).
Whereas Wigner particle creation/annihilation operators are non-local in the
sense that they can only be associated to the global algebra on all of
Minkowski spacetime\footnote{``Almost local'' operators \cite{Haag} i.e.
operators which are ``essentially'' localized in a finite region but have
``tails'' extending to spatial infinity are not considered here.}, wedge
localized one particle states can already be obtained by applying specific
operators (FPGs) associated with wedge-localized algebras to the vacuum. For
subwedge regions this is not possible in the presence of interactions, so that
regions whose causal closure is wedge-shaped are distinguished by presenting
the best compromise between the de-localizing tendency of interacting
particles and the locality principle which is historically (Faraday, Maxwell)
associated with fields. Presently the constructive use of such wedge-localized
PFGs (whose existence is guarantied by the Tomita-Takesaki modular theory) is
restricted to tempered PFGs \cite{B-B-S}, which turn out to be related to the
Zamolodchikov-Faddeev algebraic structure. The attractive aspect of this
intrinsic algebraic bootstrap formfactor approach \cite{S1}\cite{Lechner} for
mathematical physics is that the age old problem of existence of nontrivial
QFTs becomes separated for the first time from the issue of ultraviolet
behavior and instead passes to properties which refer phase-space degrees of
freedom associated with wedge-localized algebras. In d=1+1 this amounts to
check their ``nuclear modularity'' property \cite{Bu-Le}.

The non-local ''direct particle interaction'' approach of Coester and Polyzou
(section 2) has its point of departure in the extreme opposite setting of
relativistic quantum mechanics i.e. it is a framework without any vacuum
polarization structure. The only conceptually somewhat unusual aspect of this
approach is the two-step way in which the inductive (Bakamijan-Thomas)
construction of a Poincar\'{e} covariant but non-cluster-factorizing
n-particle system is followed by the enforcement of clustering via (Sokolov)
scattering equivalences. Although it was not our aim to present this direct
particle approach as a serious contender for a fundamental non-local
interactions, we think that it should be part of the intellectual preparation
for somebody who wants to study the issue of non-local interactions to be
aware that the reason why this possibility was missed in the history of the
pre-renormalization particle physics was precisely that non-locality and the
field-based formalism of QFT do not mix well.

Another advantage of contrasting the aforementioned algebraic approach to QFT
via wedge algebra generators with the direct particle interaction setting is a
possible deeper understanding of vacuum polarization as the most
characteristic property of local QFT. Since the days when Heisenberg
discovered that near the surface of spatial localization regions of
``partial'' charges there are always (infinitely big for sharp surfaces)
vacuum polarization clouds (even without interactions), and ever by since the
ancient observation Furry and Oppenheimer that interactions implemented by
Lagrangian perturbations always lead to field-states (states obtained from the
vacuum by the application of fields) whose one-particle contribution are
inexorably accompanied by vacuum polarization clouds, it became desirable to
have a more intrinsic quantitative nonperturbative understanding of these
phenomena. Important effects as the on-shell crossing property and the
localization-induced thermal effects (Hawking-Unruh radiation from causal
Rindler horizons, area proportionality of localization induced entropy etc.)
depend on a better understanding of vacuum polarization.

The non-locality behind the string-localization (section 3) has a clear
historical motivation. Far from being a mere invention, it was looming there
ever since Wigner wrote his famous article on the classification of
irreducible representations of the Poincar\'{e}-group. But whereas for all the
finite halfinteger spin representations there was a description in terms of
local Euler-Lagrange setting following the prior rules of Jordan's field
quantization, the Wigner zero mass infinite helicity representation turned out
to be outside this historically cherished framework. This is the reason why
the associated string-localized operator theory was discovered only recently;
without any classical guide one needed a good understanding of modular
localization before one could solve this problem.

The open question is whether in addition to filling a historical loophole,
string localization can lead to new physics. The hope that this turns out to
be the case is founded on the observation that there exist string-localized
massive free fields $A(x,e)$ with nice properties which, as the zero mass
infinite spin fields, live simultaneously on Minkowski spacetime (labeled by
$x)$ as well as on one dimensional lower de Sitter manifold of spacelike
directions (labeled by $e$). One nice property which emerges from a study of
their two-point function is that they are less singular in $e$-direction than
e.g. the associated pointlike fields \cite{M-S-Y2}. There is some hope that
these objects may facilitate a perturbative construction of massive
Buchholz-Fredenhagen strings whose existence is strongly suggested by methods
of algebraic QFT \cite{B-F}, but for which no convincing example has been
constructed yet.

The more popular incursions into the non-local realm are the various attempts
to formulate and apply non-local QFT. Whereas the older attempts (see
introduction) in the aftermath of renormalization theory were motivated by the
avoidance of ultraviolet divergencies and the desire to extend Lagrangian
quantization in order to incorporate particle formfactors as fundamental
concepts, the more recent approaches (section 5) aim at non-locality via
non-commutative modifications of Minkowski space. The motivation for this is
twofold, on the one hand there are good physical reasons to look at the
conceptual modifications of spacetime if one subjects the Einstein
gravitational equations to the kind of quasiclassical requirements which in
the context of the Maxwell equations in conjunction with quantum matter led
Bohr and Rosenfeld to their uncertainty relations for electromagnetic field
strength. In this spirit spacetime commutation were first derived in
\cite{D-F-R}. A more formal argument in the same direction but with a
buildt-in violation of Lorentz invariance (a kind of quantum ether) came from
string theory \cite{Sei-Wi}.

While the original purpose behind these proposals was to contribute ideas
towards quantum gravity, their present use seems to be limited to QFT in
noncommutative Minkowski spacetime. Although as non-local models of particle
physics some of these attempts are mathematically much more sophisticated than
the pedestrian old attempts via formfactor modifications of Lagrangians, they
still must pass those requirements as the cluster factorization property and
scattering theory on which the old attempts failed.

There has been a recent proposal to reconcile ideas of noncommutative geometry
with the principle of local covariance \cite{Pa-Ve}. The reason why
noncommutative geometry cannot directly be applied to particle physics is that
it is modelled on Euclidean space, and apart from very special situations one
meets in Lagrangian quantization there is no analytic bridge between Euclidean
and real spacetime. What makes this idea attractive is that not only do the
authors propose to overcome the restrictions to euclidean spectral triples,
but they seem to be the first who take the recently discovered principle of
``local covariance'' \cite{B-F-V} serious and try to adapt it to a
noncommutative context towards quantum gravity.

Acknowledgments: I am indebted to Fritz Coester and Wayne Polyzou as well as
Dorothea Bahns and Gherardo Piacitelli for an interesting correspondence. To
Karl-Henning Rehren I owe thanks for a critical reading of the manuscript.

\end{subequations}
\end{document}